# Tailoring Magnetic Anisotropy in $Cr_2Ge_2Te_6$ by Electrostatic Gating

Ivan. A. Verzhbitskiy[1,2*], Hidekazu Kurebayashi[3*], Haixia Cheng[4], Jun Zhou[1], Safe Khan[3], Yuan Ping Feng[1,2], and Goki Eda[1,2,5*]

[1]Physics department, National University of Singapore, Singapore

[2]Centre for Advanced 2D Materials, National University of Singapore, Singapore

[3]London Centre for Nanotechnology, University College London, United Kingdom

[4]Department of Physics, University of Science and Technology Beijing, 100083, China

[5]Chemistry department, National University of Singapore, Singapore

E-mail: ivan@nus.edu.sg (IV), h.kurebayashi@ucl.ac.uk (HK), g.eda@nus.edu.sg (GE)

**Electrical control of magnetism of a ferromagnetic semiconductor offers exciting prospects for future spintronic devices for processing and storing information. Here, we report observation of electrically modulated magnetic phase transition and magnetic anisotropy in thin crystal of $Cr_2Ge_2Te_6$ (CGT), a layered ferromagnetic semiconductor. We show that heavily electron-doped (∼$10^{14}$ cm$^{-2}$) CGT in an electric double-layer transistor device is found to exhibit hysteresis in magnetoresistance (MR), a clear signature of ferromagnetism, at temperatures up to above 200 K, which is significantly higher than the known Curie temperature of 61 K for an undoped material. Additionally, angle-dependent MR measurements reveal that the magnetic easy axis of this new ground state lies within the layer plane in stark contrast to the case of undoped CGT, whose easy axis points in the out-of-plane direction. We propose that significant doping promotes double-exchange mechanism mediated by free carriers, prevailing over the superexchange mechanism in the insulating state. Our findings highlight that**

**electrostatic gating of this class of materials allows not only charge flow switching but also magnetic phase switching, evidencing their potential for spintronics applications.**

Ferromagnetic semiconductors are an attractive platform to realize simultaneous electrical control of charge and spin degrees of freedom[1,2]. Recent discovery of magnetic order in atomically thin ferromagnetic semiconductors such as $CrI_3$ (Ref. [3]), $CrBr_3$ (Ref. [4]) and $Cr_2Ge_2Te_6$ (CGT)[5] motivated studies on the effect of electric field on magnetism, leading to intriguing phenomena including ferromagnetic-antiferromagnetic switching[3,6], tunable magnetization loop[7], enhanced tunnelling magnetoresistance[8,9] and magnon-assisted tunnelling[4]. These phenomena offer exciting prospects for novel spintronic devices. For practical device implementations, electrical control of magnetism at or near room temperature is desirable[1]. However, these layered ferromagnetic semiconductors exhibit Curie temperatures ($T_C$) well below 100 K (Ref. [10]). Modulation of $T_C$ by electric field has been demonstrated but with limited enhancement[6]. In this Article, we report electrical control of ferromagnetism in CGT with a giant enhancement in $T_C$ and alteration in magnetic anisotropy.

CGT is a van der Waals layered ferromagnetic semiconductor with a band gap of ~0.7 eV (Ref. [11]). It shows a ferromagnetic order below its $T_C$ of 61 K, with a small coercivity of ~3.4 mT (Ref. [12]). The ferromagnetic order in CGT is governed by the intra-layer superexchange[13] coupling through a Cr-Te-Cr bond with ~90° bond angle (Fig. 1a), which is well explained by Goodenough-Kanamori rule[14,15], and a weak interlayer ferromagnetic coupling[5]. CGT exhibits a strong magnetic anisotropy with out-of-plane easy axis[12]. The magnetic order persists when the material is thinned down to bilayer limit despite with reduced $T_C$ (Ref. [5]). Recently, electrostatic control of magnetism in CGT in a field-effect device geometry has shown remarkable enhancement of saturation magnetization but with no changes in $T_C$ (Ref. [7]). This strongly contrasts with the case of electric field effect in $Fe_3GeTe_2$ (FGT), a metallic analogue of CGT, where $T_C$ was increased by 200 K by electrochemical doping, which induced high electron density of ~$10^{14}$ cm$^{-2}$ (Ref. [16]). This contrasting behaviour of CGT and FGT suggest the importance of high-density doping. Recent report on significantly enhanced $T_C$ of a hybrid superlattice of CGT and tetrabutyl ammonium (TBA) further suggests strong doping dependence of this material[17]. However, the effect of carriers on magnetic order in CGT in the high density regime is largely unexplored.

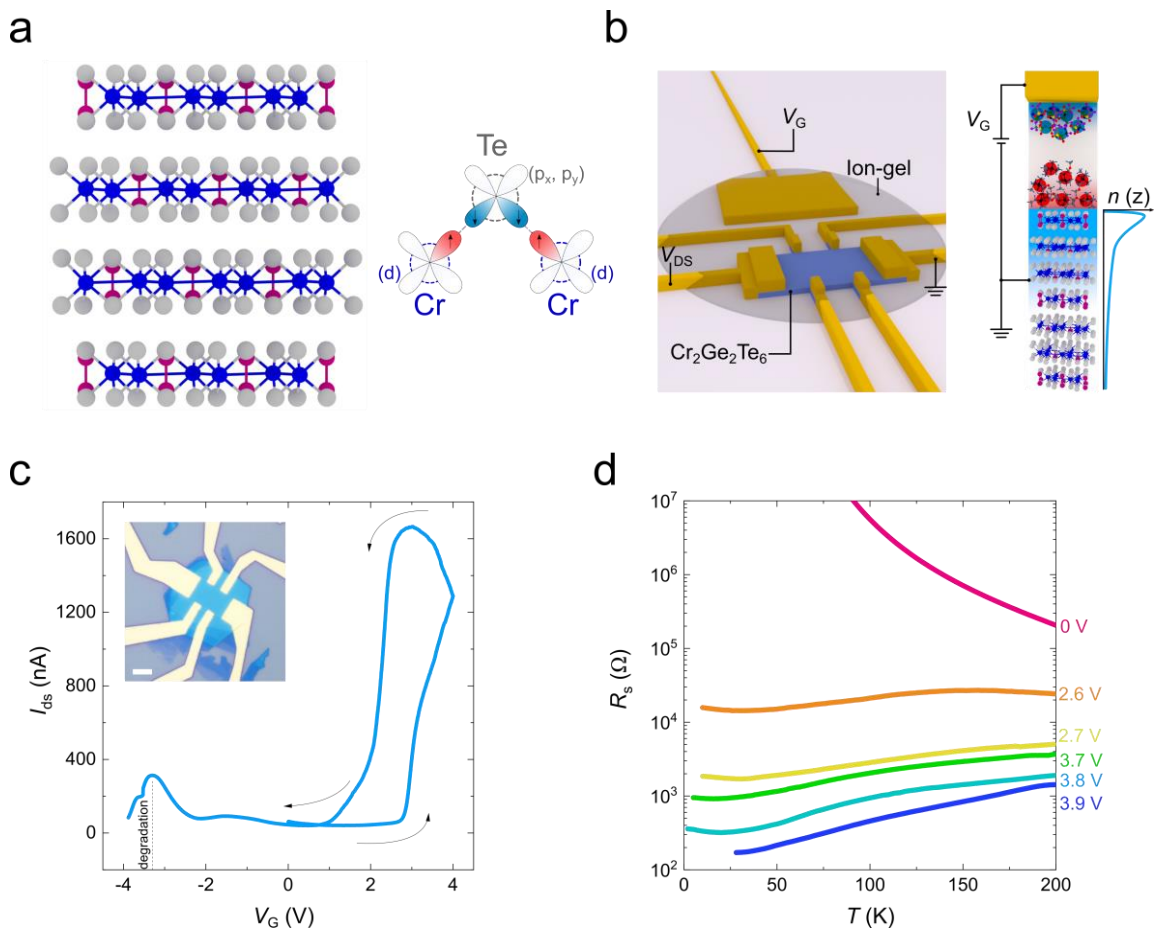

**Figure 1. Characterization of CGT.** (a) Ball-and-stick model of the CGT crystal structure (left) and a schematic illustration of the superexchange coupling established via virtual hopping of the electrons through the Cr-Te-Cr 90° bonding (right); (b) Schematic of an EDLT device based on CGT (left) and carrier density distribution along the thickness direction (right). Application of the positive gate bias $V_G$ drives cations in ion gel towards the surface of CGT, thus forming an electric double-layer. Most of potential drop occurs at the EDLT interface. The accumulation layer in CGT occupies only a few top-most layers; (c) Typical ambipolar transfer curve of the device at $V_{DS}$ = 25 mV and $T$ = 250 K; an optical micrograph of the ~20 nm thin CGT flake with Pd/Au contacts is shown in the inset (scale bar is 5 µm); (d) $R_s$-$T$ curves at different $V_G$. The sheet resistance, is reduced by an order of magnitude at $T$ = 200 K when $V_G$ was increased from 2.6 V to 3.9 V. At these gate biases, CGT shows metallic behaviour in stark contrast with the undoped ($V_G$ = 0) regime.

To achieve high doping levels in CGT, we utilized the EDLT geometry with a polymer gel[18] based on ionic liquid (DEME-TFSI) that provides a high-density carrier accumulation at the gel/semiconductor interface due to its ultra large capacitance (~10 μF/cm$^2$)[19]. This electrolyte was previously used to induce interfacial superconductivity in transition metal dichalcogenides[20, 21] and magnetism in cobalt-doped titanium dioxide[22]. Due to strong screening effects, the induced carriers are confined in a few topmost layers (Fig. 1b), thus effectively creating a 2D electron gas (Ref. [19], Supplementary Section 4). A multi-probe device with a mechanically exfoliated CGT flake (~ 20 nm) was covered with an ion gel as schematically illustrated in Fig. 1b. Figure 1c shows the transfer characteristic of a typical device at 250 K. The ambipolar character of CGT is evident with electron current increasing rapidly with gate bias ($V_G$) above ~3 V and hole current emerging below ~-2V. Decreasing $V_G$ below ~-3.5 V led to irreversible degradation of the material most likely due to electrochemical reaction. On the other hand, the device was stable in the electron-doped regime. For $V_G$ between 2.6 and 4.0 V, the sheet resistance, $R_s$, of the device decreases with decreasing temperature, indicating the metallic character of the heavily doped CGT (Fig. 1d). This is in clear contrast to undoped ($V_G$ = 0 V) and weakly doped material,[7, 23] where the resistance quickly diverges with decreasing temperature, indicating insulating behaviour. We conducted *in-situ* Raman spectroscopy to verify that the onset of the high conductivity regime is not accompanied by any structural changes (Supplementary Section 2). Thus, the observed metal-insulator transition[24] reflects the effective filling of the band edge states. From Hall effect measurements, we estimate the electron density to be ~ 4×10$^{14}$ cm$^{-2}$ for $V_G$ = 3.9 V (Supplementary Section 3).

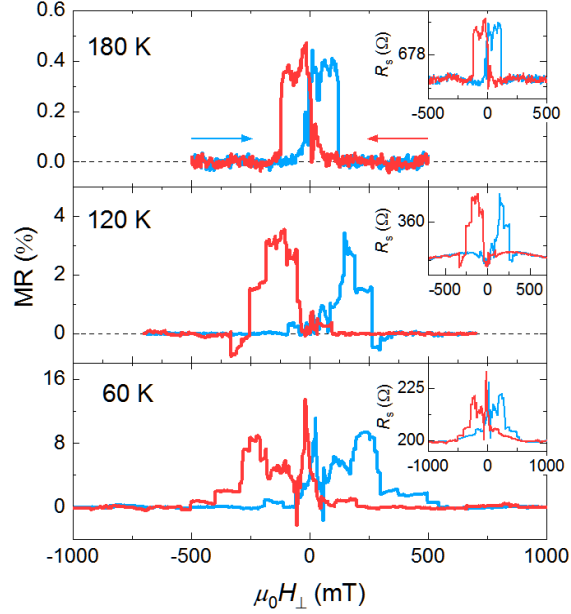

**Figure 2. MR hysteresis**. Magnetoresistance ($MR = (R_s(H) - R_s(0))/R_s(0)$) curves for $T$ = 180 K (top), 120 K (middle) and 60 K (bottom) and $V_G$ = 3.9 V (device #1). The background is removed for clarity. The magnetic field is applied in the out-of-plane direction. Unprocessed data are shown in the insets.

Here, we monitor the longitudinal resistance change as a function of applied magnetic field to probe the magnetic state of the material. Figure 2 shows the MR curves at 60, 120, and 180 K with magnetic field applied normal to the plane. Remarkably, the MR exhibits a clear hysteresis with abrupt changes in resistance at low fields. This MR hysteresis, similar to the recently observed hysteresis in magnetic PdSe$_2$ (Ref. [25]), strongly indicates the presence of spontaneous magnetization and magnetic anisotropies (Supplementary Section 5). Most devices consistently exhibited MR hysteresis at these temperatures when sufficiently large doping was achieved, indicating that doping induces magnetic order well above $T_C$ of undoped CGT. The microscopic origin of the complex resistance changes is unclear, but a series of steps suggests the presence and depinning behaviour of magnetic domains with sizes smaller than the device channel. We further conducted angle-dependent MR to identify the origin of resistance changes (Fig. S5). However, the results could not be explained by the conventional model, suggesting that multiple mechanisms are at play. Nevertheless, the abrupt resistance steps are highly reproducible (Fig. S6) and occur at the same magnetic fields, indicating that the observed MR hysteresis reflects the magnetization switching mechanism of the material.

Remarkably, we found that not only $T_C$ but also the magnetic anisotropy of CGT is altered by the application of electric field. Figure 3a shows the MR hysteresis measured with different tilt angles $\gamma$ between the applied magnetic field and the plane of CGT crystal. It is evident that it requires a significantly larger magnetic field to complete magnetisation switching for out-of-plane fields (See Fig. S6 for temperature dependence of in-plane MR). To discuss this quantitatively, we define the effective saturation field $H_{sat}$ to be the field where the magnetisation switching completes, *i.e.* the field at which the hysteresis disappears. Figure 3b shows that $H_{sat}$ rises sharply as the tilt angle approaches 90 degrees. This is a typical characteristic of magnetisation switching associated with domain walls [26, 27] that explains magnetic switching behaviours in various systems[28, 29, 30]. It provides the angle dependence of the depinning field $H_{dep}(\gamma)$ of domain walls from the easy and hard axes as $H_{dep}(\gamma) = \frac{H_{dep}^0}{\cos \gamma}$ where $H_{dep}^0$ is the depinning field along the easy axis. This model shows an excellent fit to our experimental results, indicating that the dominant switching mechanism is the domain-wall motion. Note that the small deviation of the peak from 90° ($\Delta\gamma = 2.5°$) is due to backlash of the rotation probe in our experimental setup. This model further verifies that the magnetic easy axis is along the in-plane directions for heavily doped CGT, in contrast to the out-of-plane easy axis of undoped CGT. The observed magnetic anisotropy is similar to that of TBA/CGT superlattice[16]. We highlight that the induced anisotropy in our device is not due to ion-intercalation but due to surface doping.

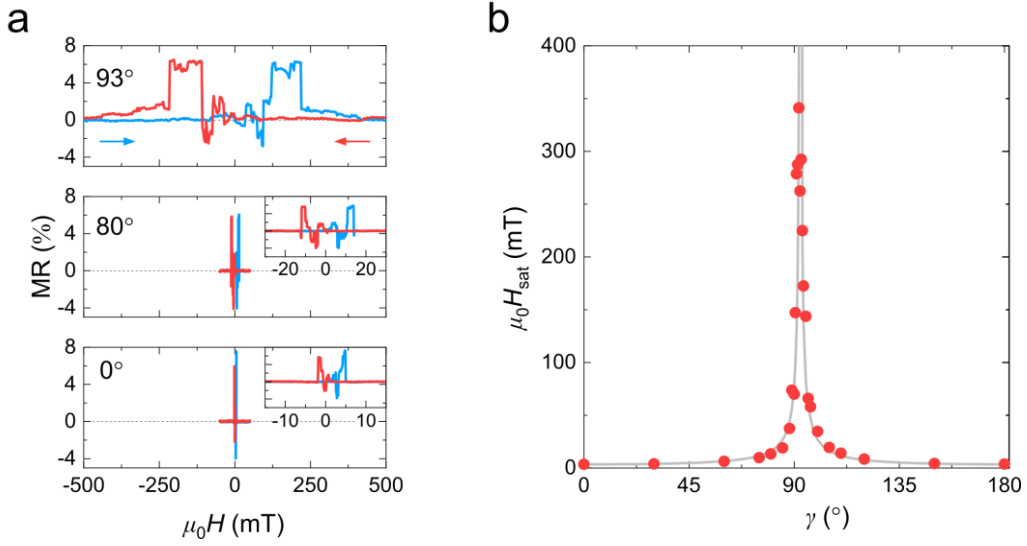

**Figure 3. Angle dependence of magnetoresistance.** (a) MR curves at different angles $\gamma$ between 2D plane and the magnetic field. Insets in middle and bottom panels highlight the persistent hysteresis in the small field regime; (b) Extracted saturation field ($H_{sat}$) as a function of $\gamma$ where solid curve represents the fit based on the domain wall depinnig model.

To investigate the role of carrier density in inducing ferromagnetic order in CGT, we conducted temperature dependent MR measurements at different gate voltages through multiple cooling cycles. Figures 4a-c show colour plots of $\Delta\text{MR} = |\text{MR}^\uparrow - \text{MR}^\downarrow|$ as the function of the applied out-of-plane magnetic field and temperature for three different gate bias conditions of a single device. Here, $\text{MR}^\uparrow$ ($\text{MR}^\downarrow$) is the magnetoresistance measured from negative to positive (positive to negative) fields. Transition from ferromagnetic to paramagnetic phase is evident from the disappearance of $\Delta\text{MR}$ at higher temperatures. $T_C$ for $V_G = 2.6$ V condition is higher than that of pristine CGT by a factor of two, and an increase in $V_G$ from 2.6 to 3.9 V further enhances $T_C$ by nearly 80 K. We conducted gate bias dependent measurements on two different devices and both were found to exhibit the same trend (Fig. S8).

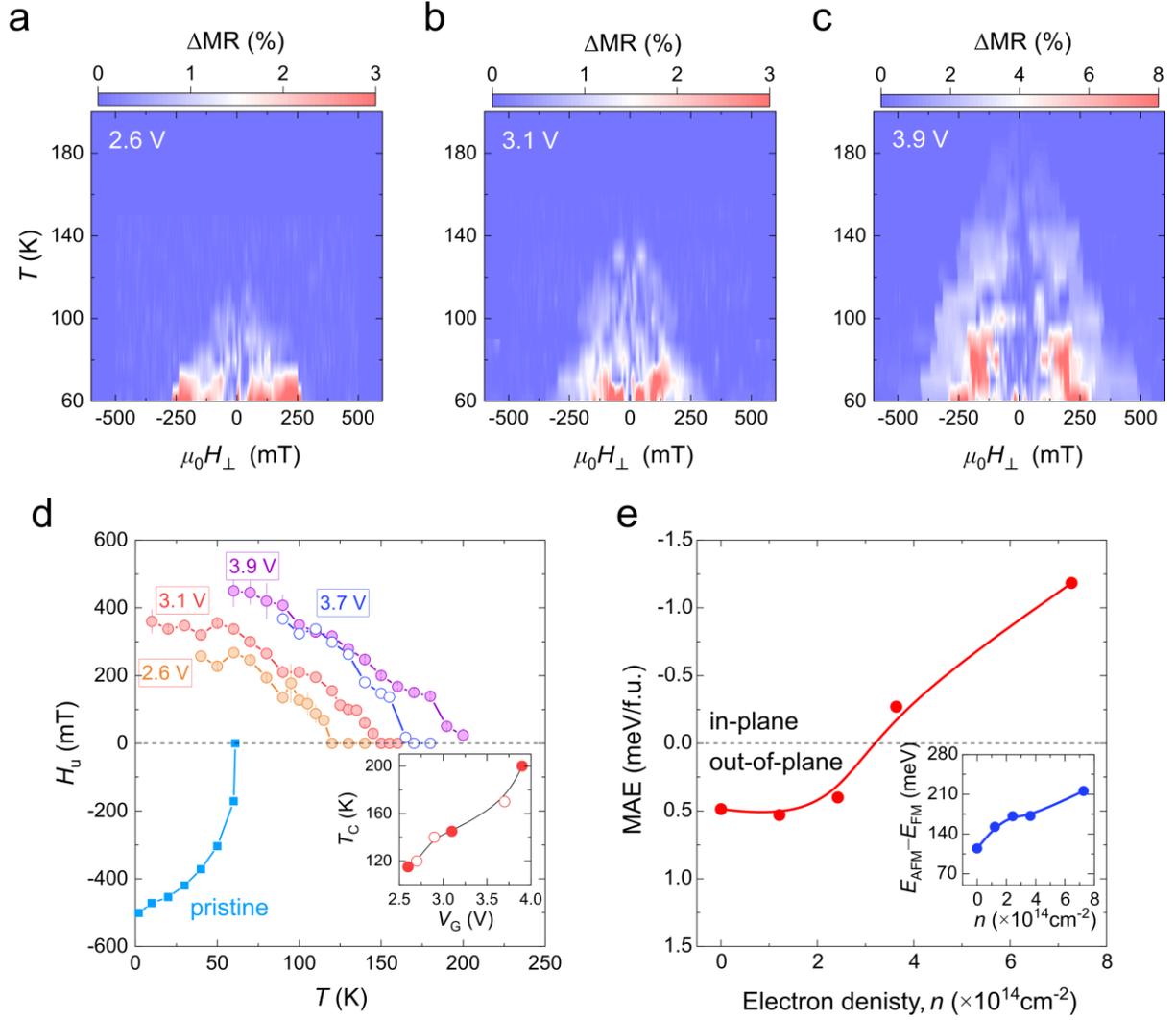

**Figure 4. Carrier density dependence of ferromagnetism.** Colour plots of $\Delta MR = |MR^\uparrow - MR^\downarrow|$ for (a) $V_G$ = 2.6 V, (b) 3.1 V and (c) 3.9 V, where $MR^\uparrow$ ($MR^\downarrow$) is the magnetoresistance curve measured with increasing (decreasing) field; (d) $H_u$, defined as $(H_{sat}^\perp - H_{sat}^\parallel)$, as a function of the gate bias. Square symbols represent values measured for a pristine bulk sample (data are taken from Ref. [31]), while the round filled (clear) symbols depict the data from the EDLT device #1 (device #2). Inset highlights the dependence of $T_C$ on $V_G$. (e) Calculated MAE as a function of electron density. Change of $E_{AFM}$-$E_{FM}$ at different doping densities is shown in the inset. The solid lines are guide to the eye.

We further conducted in-plane field MR to extract the uniaxial magnetic anisotropy fields $H_u (= H_{sat}^\perp - H_{sat}^\parallel)$ for different $V_G$ as shown in Figure 4d. For undoped CGT, $H_u$ was obtained by ferromagnetic resonance experiments[31]. Note that $H_u < 0$ and $H_u > 0$ correspond to the magnetic easy axis lying in the out-of-plane and in-plane directions, respectively. For all gate biases, $|H_u|$ decreases with increasing temperature and approaches zero. We estimate $T_C$ from the intercept with the $H_u = 0$ line. The trend for undoped bulk CGT is the same except that the sign of $H_u$ is opposite to that of the doped CGT. The increasing trend of $T_C$ with $V_G$ is consistent for the two devices (Fig. 4d, inset). The switching of the magnetic easy axis indicates the sign change in the magnetic anisotropy energy (MAE). Figure 4e shows MAE of CGT for different carrier densities obtained by density functional theory (DFT) calculations. For pristine undoped CGT, the positive MAE of 0.48 meV per formula unit (f.u.) indicates that out-of-plane spin orientation is more stable than the in-plane orientation, in agreement with the previous study[32]. With increasing electron density, MAE decreases, continuing the previously calculated trends in the low density regime[7], and becomes negative for electron density above $3\times10^{14}$ cm$^{-2}$. This general trend is in accordance with our experimental observation of easy axis switching for heavily doped CGT.

From DFT calculations, we further obtained the difference between the total energy of ferromagnetic and anti-ferromagnetic spin states ($E_{\text{AFM}} - E_{\text{FM}}$). This quantity is proportional to the magnitude of the exchange interaction energy since $E_{(\text{A})\text{FM}} = E_0 - (+)E_{\text{ex}}$ where $E_0$ is the non-magnetic free energy component and $E_{\text{ex}}$ is the exchange energy. The calculated $E_{\text{AFM}} - E_{\text{FM}}$ as a function of electron density (Fig. 4e, inset) reveals the increasing trend of the total exchange energy, implying that $T_\text{C}$ increases with doping, consistent with our experimental observations. The relationship between $E_{\text{AFM}} - E_{\text{FM}}$ and doping density is in a linear fashion to the first order, and therefore explains that no distinct $T_\text{C}$ enhancement was observed in the recent study where doping density was of the order of $10^{12}$ cm$^{-2}$ (Ref. [7]).

We now consider the origin of doping-induced magnetic order in CGT. The large carrier densities achieved in our samples suggest that the exchange interaction between magnetic Cr ions could depart from the superexchange mechanism, which requires the well-isolated $Cr^{3+}$ states, *i.e.* electrically insulating states. When some of the $Cr^{3+}$ ions are replaced with $Cr^{2+}$ ions by local electron doping (Fig. S9), the double-exchange interaction mechanism, where the spin of electrons is preserved during hopping across $Cr^{3+}$-Te-$Cr^{2+}$ links, can further stabilize the ferromagnetic order[33]. This carrier-mediated indirect exchange mechanism explains the emergence of ferromagnetism in several non-itinerant magnetic systems by chemically-induced doping[34, 35]. Electrostatically-induced carriers in CGT can act in a similar manner to cause the enhancement of $T_C$ by this mechanism. We use the electron transport parameters to model the hopping rate through $Cr^{3+}$-Te-$Cr^{2+}$ links and evaluate the energy scale of the exchange interaction (Supplementary Section 6). This phenomenological model estimates the magnitude of the exchange interaction $E_{ex}$ to be ~175 meV for $V_G = 3.9$ V, which is of the same order of magnitude as $E_{AFM} - E_{FM}$ obtained by DFT calculations (Fig. 4e, inset). Although this model only provides us with a crude estimate of $E_{ex}$, the fair agreement of the energy scale supports the role of the double-exchange mechanism.

In summary, we have demonstrated the emergence of ferromagnetic order with significantly enhanced $T_C$ in CGT crystals in an EDLT geometry. Heavy electron doping not only enhances $T_C$ but also changes the sign of the magnetic anisotropy energy, resulting in the change of the magnetic easy axis from out-of-plane to in-plane. Our analysis suggests that the carrier-mediated indirect exchange mechanism prevails over the superexchange mechanism upon doping. We envision that this approach is applicable to other insulating ferromagnets or even non-magnetic systems in a vicinity of magnetic order. Further studies will underpin the detailed role of electric field in controlling magnetism by the route presented in our study. Our findings show that layered ferromagnetic semiconductors are an exciting platform for investigating electrically tunable exchange interactions and exploring novel spintronic device concepts with electric field.

## Methods

Methods and additional references are available at the Supplementary Information.

## Acknowledgement

G.E. acknowledges the Singapore National Research Foundation for funding the research under medium-sized centre programme. G.E. also acknowledges support from the Ministry of Education (MOE), Singapore, under AcRF Tier 2 (MOE2017-T2-1-134). F.Y.P. would like to thank Shen Lei for fruitful discussion.

## Author contributions

I.V. and G.E. conceived the idea of the experiments. I.V. synthesized the CGT crystals and performed the transport measurements. H.C and J.Z. conducted the first-principles calculations with input from F.Y.P. Data analysis and interpretations were carried out by H.K., I.V., G.E. and all other co-authors. I.V., G.E., and H.K. wrote the manuscript with input from other co-authors.

## Competing interests

The authors declare no competing interests.

## Data availability

The data that support the plots within this paper and other findings of this study are available from the corresponding authors upon reasonable request.